\documentclass{article}
\usepackage{subfloat}
\usepackage{lscape,epsfig}
\usepackage{graphicx}
\usepackage{amssymb}
\usepackage{amsmath}
\usepackage{natbib}
\textheight=22cm \DeclareSymbolFont{ppa}{OT1}{ppl}{m}{it}
\DeclareMathSymbol{\vv}{\mathalpha}{ppa}{'166}

minus 2.3mu \thickmuskip = 2.6mu plus 2mu minus 2.6mu

\begin{document}

\newcommand{\dd}{\,{\rm d}}
\newcommand{\ie}{{\it i.e.},\,}
\newcommand{\etal}{{\it $et$ $al$.\ }}
\newcommand{\eg}{{\it e.g.},\,}
\newcommand{\cf}{{\it cf.\ }}
\newcommand{\vs}{{\it vs.\ }}
\newcommand{\zdot}{\makebox[0pt][l]{.}}
\newcommand{\up}[1]{\ifmmode^{\rm #1}\else$^{\rm #1}$\fi}
\newcommand{\dn}[1]{\ifmmode_{\rm #1}\else$_{\rm #1}$\fi}
\newcommand{\upd}{\up{d}}
\newcommand{\uph}{\up{h}}
\newcommand{\upm}{\up{m}}
\newcommand{\ups}{\up{s}}
\newcommand{\arcd}{\ifmmode^{\circ}\else$^{\circ}$\fi}
\newcommand{\arcm}{\ifmmode{'}\else$'$\fi}
\newcommand{\arcs}{\ifmmode{''}\else$''$\fi}
\newcommand{\MS}{{\rm M}\ifmmode_{\odot}\else$_{\odot}$\fi}
\newcommand{\RS}{{\rm R}\ifmmode_{\odot}\else$_{\odot}$\fi}
\newcommand{\LS}{{\rm L}\ifmmode_{\odot}\else$_{\odot}$\fi}

\newcommand{\Abstract}[2]{{\footnotesize\begin{center}ABSTRACT\end{center}
\vspace{1mm}\par#1\par \noindent {~}{\it #2}}}

\newcommand{\TabCap}[2]{\begin{center}\parbox[t]{#1}{\begin{center}
  \small {\spaceskip 2pt plus 1pt minus 1pt T a b l e}
  \refstepcounter{table}\thetable \\[2mm]
  \footnotesize #2 \end{center}}\end{center}}

\newcommand{\TableSep}[2]{\begin{table}[p]\vspace{#1}
\TabCap{#2}\end{table}}

\newcommand{\FigCap}[1]{\footnotesize\par\noindent Fig.\  %
  \refstepcounter{figure}\thefigure. #1\par}

\newcommand{\TableFont}{\footnotesize}
\newcommand{\TableFontIt}{\ttit}
\newcommand{\SetTableFont}[1]{\renewcommand{\TableFont}{#1}}
\newcommand{\MakeTable}[4]{\begin{table}[htb]\TabCap{#2}{#3}
  \begin{center} \TableFont \begin{tabular}{#1} #4
  \end{tabular}\end{center}\end{table}}

\newcommand{\MakeTableSep}[4]{\begin{table}[p]\TabCap{#2}{#3}
  \begin{center} \TableFont \begin{tabular}{#1} #4
  \end{tabular}\end{center}\end{table}}

\newenvironment{references}%
{ \footnotesize \frenchspacing
\renewcommand{\thesection}{}
\renewcommand{\in}{{\rm in }}
\renewcommand{\AA}{Astron.\ Astrophys.}
\newcommand{\AAS}{Astron.~Astrophys.~Suppl.~Ser.}
\newcommand{\ApJ}{Astrophys.\ J.}
\newcommand{\ApJS}{Astrophys.\ J.~Suppl.~Ser.}
\newcommand{\ApJL}{Astrophys.\ J.~Letters}
\newcommand{\AJ}{Astron.\ J.}
\newcommand{\IBVS}{IBVS}
\newcommand{\PASP}{P.A.S.P.}
\newcommand{\Acta}{Acta Astron.}
\newcommand{\MNRAS}{MNRAS}
\renewcommand{\and}{{\rm and }}
\section{{\rm REFERENCES}}
\sloppy \hyphenpenalty10000
\begin{list}{}{\leftmargin1cm\listparindent-1cm
\itemindent\listparindent\parsep0pt\itemsep0pt}}%
{\end{list}\vspace{2mm}}

\def\TYLDA{~}
\newlength{\DW}
\settowidth{\DW}{0}
\newcommand{\dw}{\hspace{\DW}}

\newcommand{\refitem}[5]{\item[]{#1} #2%
\def\REFARG{#3}\ifx\REFARG\TYLDA\else, {\it#3}\fi
\def\REFARG{#4}\ifx\REFARG\TYLDA\else, {\bf#4}\fi
\def\REFARG{#5}\ifx\REFARG\TYLDA\else, {#5}\fi.}

\newcommand{\Section}[1]{\section{#1}}
\newcommand{\Subsection}[1]{\subsection{#1}}
\newcommand{\Acknow}[1]{\par\vspace{5mm}{\bf Acknowledgements.} #1}
\pagestyle{myheadings}

\newfont{\bb}{ptmbi8t at 12pt}
\newcommand{\xrule}{\rule{0pt}{2.5ex}}
\newcommand{\xxrule}{\rule[-1.8ex]{0pt}{4.5ex}}
\def\thefootnote{\fnsymbol{footnote}}

\begin{center}
{\Large\bf
 The Clusters AgeS Experiment (CASE).  \\
 Variable stars in the field of  
 the globular cluster NGC 6362\footnote{Based
 on data obtained with du Pont and Swope telescopes at Las
 Campanas Observatory.}}
 \vskip1cm
  {\large
      ~~J.~~K~a~l~u~z~n~y$^1$,
      ~~I.~B.~~T~h~o~m~p~s~o~n$^2$,
      ~~M.~~R~o~z~y~c~z~k~a$^1$,
      ~~W.~~P~y~c~h$^1$~~and~~W.~~Narloch$^1$
   }
  \vskip3mm
{ $^1$Nicolaus Copernicus Astronomical Center, ul. Bartycka 18, 00-716 Warsaw, Poland\\
     e-mail: (jka, mnr, pych, wnarloch)@camk.edu.pl\\
  $^2$The Observatories of the Carnegie Institution of Washington, 813 Santa Barbara
      Street, Pasadena, CA 91101, USA\\
     e-mail: ian@obs.carnegiescience.edu}
\end{center}

\vspace*{7pt}
\Abstract 
{The field of the globular cluster NGC~6362 was monitored between 1995 
and 2009 in a search for variable stars. $BV$ light curves were obtained 
for 69 periodic variables including 34 known RR Lyr stars, 10 known 
objects of other types and 25 newly detected variables. 
Among the latter we identified 18 proper-motion 
members of the cluster: seven detached eclipsing binaries (DEBs), six SX Phe 
stars, two W UMa binaries, two spotted red giants, and a very interesting 
eclipsing binary composed of two red giants - the first example of such a 
system found in a globular cluster. Five of the DEBs are located at the 
turnoff region, and the remaining two are redward of the lower main sequence. 
Eighty-four objects from the central $9\times9$ arcmin$^2$ of the cluster
were found in the region of cluster blue stragglers. Of these 70 are proper 
motion (PM) members of NGC 6362 (including all SX Phe and two W UMa stars), 
and five are field stars. The remaining nine objects lacking PM information are 
located at the very core of the cluster, and as such they are likely genuine 
blue stragglers. 
}
{globular clusters: individual (NGC 6362) -- stars: variables -- 
stars: SX Phe -- blue stragglers -- binaries: eclipsing
}

\Section{Introduction} 
\label{sec:intro}
NGC~6362 is a nearby ($(m-M)_{V}=14.68$ mag) globular cluster located at 
a rather high galactic latitude ($b=-17.6$~deg) in a field of low reddening 
with $E(B-V)=0.09$ (Harris 1996; 2010 edition).
These properties together with a low concentration make it
an attractive target for detailed studies with ground based telescopes.
The photometric survey presented here is a part of the CASE project (Kaluzny 
et al. 2005) conducted with telescopes of the Las Campanas Observatory.

Early pre-CCD searches for variable stars in the cluster
were summarized by Clement et al. (2001). They resulted in the detection
of 31 RR Lyr stars. Based on CCD-data, Mazur et al. (1999) found 19 new 
variables (among them five RR Lyr stars, four SX Phe stars and eight 
eclipsing binaries). Periods were  established only for short period variables 
with $P<1$ d. Subsequently, Olech et al. (2001) performed an analysis of light 
curves for 35 RR Lyr stars from the cluster. Three of these turned out to be 
non-radial pulsators.

In this contribution we present results of an extended survey
conducted at Las Campanas Observatory between 1995 and 2009.
Section 2 contains a report on the observations and explains the
methods used to calibrate the photometry. The detected variables
are presented and discussed in Section 3. The paper is summarized 
in Section 4.
\section{Observations}
\label{sec:obs}
This paper is based on two sets of images collected at Las
Campanas Observatory. The first  was obtained using the 2.5-m
du Pont telescope and the $2048\times 2048$ TEK5 CCD camera with
a field of view of 8.84 arcmin on a side at a scale of 0.259
arcsec/pixel. Observations were conducted on 45 nights from April 
21, 1995 to September 26, 2009; always with the same set of 
filters. For the analysis, we used 1748 $V$-images with  seeing 
ranging from 0.79 to 2.48 arcsec, and 334 $B$-images with seeing 
ranging from 0.86 to 2.15 arcsec. The median value of the seeing was 
1.32 and 1.38 arcsec for $V$ and $B$, respectively. The second set 
of images was obtained with the 1.0-m Swope telescope and the $2048\times 
3150$ SITE3 CCD camera.
The field of view was $14.8\times 22.8$ arcmin$^2$ at a scale of 0.435
arcsec/pixel. About 30\% of the images were taken with a subraster
providing a field of view of $14.8\times 14.8$ arcmin$^2$.
Observations were conducted on 103 nights from July 08, 1999 to September 
09, 2009. The same filters were used for all observations.
For the analysis, we used 3200 $V$-images with seeing ranging from 0.80 
to 2.25 arcsec and 558 $B$-images with seeing ranging from 0.94 to 2.17 
arcsec. The median value of the seeing was 1.49 and 1.55 arcsec for $V$ 
and $B$, respectively. 

\begin{figure}
   \centerline{\includegraphics[width=0.95\textwidth,
               bb = 23 239 492 495, clip]{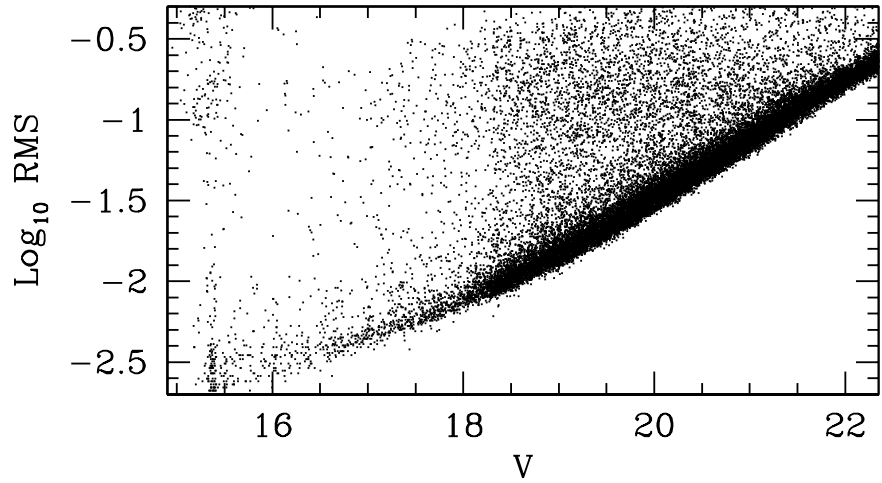}}
   \caption{ Standard deviation vs. average $V$ magnitude for
    light curves of stars from the NGC~6362 field. Light curves
    are based on images from the du Pont telescope.
    \label{fig:rms}}
\end{figure}

The photometry was measured using the image subtraction technique.
For the du Pont data we used a modified version of the ISIS V2.1 package 
(Allard  2000). For the images collected with the Swope telescope the DIAPL 
package\footnote{http://users.camk.edu.pl/pych/DIAPL/index.html} was used.
For each set and each filter, a reference image was constructed
by combining several high quality frames. Daophot, Allstar and Daogrow codes 
(Stetson 1987, 1990) were used to extract the profile photometry, and to
derive aperture corrections for the reference images.
We also extracted  profile photometry for individual images from
the du Pont telescope. This allowed us to obtain useful measurements for
stars whose profiles were overexposed on the reference images. Moreover,
this profile photometry enabled an unambiguous identification of variable
stars in crowded fields, which is sometimes problematic when  image 
subtraction only is used. We used a list of stars based on du Pont photometry
when reducing the Swope data for the central part of the cluster. 
The better spatial resolution of du Pont images allowed us to 
resolve numerous blends on images obtained with the smaller telescope.
The accuracy of the du~Pont photometry is illustrated in Fig. \ref{fig:rms},
in which the standard deviation of the photometric measurements is plotted 
as a function of the average magnitude in~$V$. 
\subsection{Calibration}

\begin{figure}
   \centerline{\includegraphics[width=0.95\textwidth,
               bb = 56 57 562 380, clip]{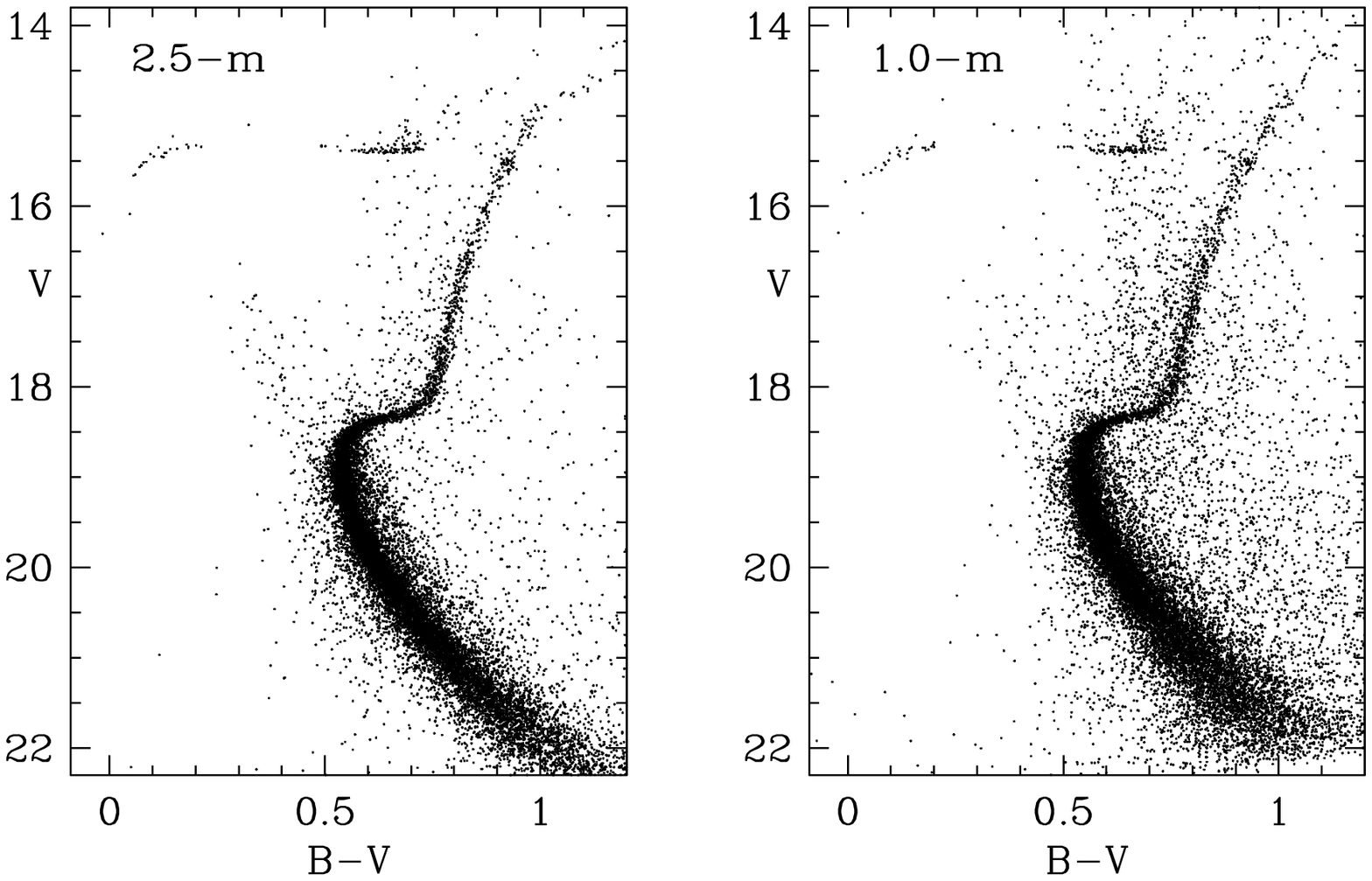}}
   \caption{Color-magnitude diagrams for NGC~6362 based on the data 
    from the du Pont telescope (left) and Swope telescope (right)
    \label{fig:cmd}}
\end{figure}

The photometry collected with du Pont telescope was transformed to the
standard $BV$ system based on observations of stars from
Landolt fields (Landolt 1992). On the night of May 3, 2001 we
observed 19 stars from four such fields, each of them observed twice.
These data were used to find the coefficients of linear transformation 
from the instrumental system to the standard one. Residual differences 
between the standard and recovered magnitudes and colors amounted to 0.009, 
0.008 and 0.010 mag for $V$, $B$ and $B-V$, respectively. 
The residuals did not show any systematic dependence on the color index.
Transformations for  photometry obtained with the Swope telescope were
based on the calibrated data from du Pont telescope.
The linear transformations proved to be entirely adequate.
Fig. \ref{fig:cmd} shows the color-magnitude diagram (CMD) of the 
observed fields (RR Lyr stars are not shown). This figure shows in particular
the wide range of stellar population examined for variability. Clearly
visible, especially in the left panel, is a rich group of
blue stragglers. The  relative contamination of the cluster 
by field interlopers increases with the increasing field of 
view. Fig.~\ref{fig:cmd} is based on reference images. Stars 
with large formal errors in $V$ and $B-V$ or those at large distances
from the cluster center are not shown. Full sets of photometry 
can be downloaded from the CASE archive.\footnote{http://case.camk.edu.pl/}. 
\subsection{Search for variables}
The search for variable stars was conducted using AOV and AOVTRANS
algorithms implemented in the TATRY code (Schwarzenberg-Czerny
1996, Schwar\-zenberg-Czerny \& Beaulieu 2006). We examined the
du Pont light curves of 13299 stars with $V<22.0$ and the Swope 
light curves of 18754 stars with $V<20.75$. The limits of detectable 
variability depended on the accuracy of photometric measurements, which
for the du Pont data decreased from 3~mmag at $V = 16$~mag to 30~mmag at $V=20$~mag 
and 100~mmag at $V=22$~mag (Fig.~\ref{fig:rms}). For the Swope data it decreased 
from 3.5~mmag at $V=15$~mag through 32~mmag at $V=19.25$~ mag to 100~mmag at 
$V=20.75$ mag.

\section{Variables}

We detected a total of 69 certain variables of which 48 have photometry
from both telescopes. Among these were 34 known RR Lyr stars, whose 
light curves will be analyzed in a separate paper (Moskalik et al., in preparation).
The basic properties of the remaining variables are listed in Table 1
together with their equatorial coordinates. The coordinates are given in the  
UCAC4 system (Zacharias et al. 2013 ) and are accurate to about 0.2 arcsec.
Variables V53-V77 are new detections; their finding charts 
are presented in Fig. 3. 

The $V$ magnitudes  listed in Table 1 correspond to the maximum light in the 
case of eclipsing binaries, while for the remaining variables average magnitudes
are provided. For each variable the $B-V$ color is given, followed by the amplitude 
in the $V$-band. Periods of variability were found for all stars. However, the 
light curves of some objects classified as "spotted" are not coherent and show 
phase shifts from season to season. In such cases, we give periods obtained for 
the indicated season(s). The last column of Table 1 gives the membership status based on 
proper motions taken  from Zloczewski et al. (2012) and Narloch 
et al. (in preparation). Phased light curves of the variables from Table 1, ordered 
according to the type of variability, are presented in Fig. 4. 

A CMD of the cluster with the locations of the variables is shown in Fig. 5.  
This CMD is based on the du Pont photometry and includes only stars selected by Zloczewski 
et al (2012) as proper motion members of NGC 6362. Variables which, based on their 
proper motions, likely belong to the cluster, are marked in red; the remaining ones - in blue.
The location of each variable coincides with the center of the corresponding label. 

\subsection{Eclipsing binaries}
We detected a total of 12 detached eclipsing binaries.
Of these, seven are proper motion members of the cluster.
Binaries V40, V41 V62, V65 and V71 are located at the cluster turnoff region. Their orbital periods 
range from 5.2 d for V40 to 17.9 d for V41. These systems are potentially 
interesting targets for a detailed follow up study aiming at  
the determination of their absolute parameters, and the age and distance of the cluster. 
We  are presently 
conducting such an analysis for V40 and V41 (Kaluzny et al., in preparation).
The variability of these two systems was first reported by 
Mazur et al. (1999). Based on the present photometry, we determined 
precise ephemerides and obtained complete light curves for both
binaries. It is remarkable that V40 has an eccentric orbit despite 
a relatively short period. At an age of about 12 Gyr (Harris 1996, 2010 edition),
its orbit should have been fully circularized  by the tidal friction 
mechanism (Mazeh 2008; Mathieu et al. 2004). There is no evidence for the presence 
of a third body in this system. We speculate that the orbit of
V40 was distorted during the last 
few Gyr as the result of a  close encounter.

The light curve of V62 shows two partial eclipses of similar depth. As a result the 
photometric solution is likely to be degenerate, allowing for a broad range of
relative radii. Such a degeneracy may be overcome by the determination of the 
light ratio from  spectra. However, this approach would be problematic given the 
faintness of the system. Similar comments apply to V71.  Much more 
promising in this respect is the binary V65. Our phased light curve of this system 
shows one eclipse with a depth of 0.45 mag. We did not cover
the bottom of the second eclipse, but the data suggest that both eclipses are close
to totality. If this is the case then the analysis of full light curves would be 
straightforward.

Another variable deserving a detailed study is V49. 
This is an eclipsing PM member of the cluster located 0.5 mag above the 
red giant branch at $V=15.0$. The orbital period of 32 days,  
a $\beta$~Lyr type light curve and the location on the CMD indicate that V49 is composed 
of two red giants forming a close binary. Figure~\ref{fig:V49} shows phased 
$V$ and $B-V$ light curves. The secondary 
eclipse is total, with the ingress starting at a phase of about 0.46. The 
duration of totality is $\sim0.06P$. In the secondary minimum 
the system becomes markedly redder, indicating that the smaller and hotter
component is eclipsed. The depth of the secondary eclipse varies by $\sim0.04$ 
mag; presumambly due to a spot activity on at least one of the components.
These variations and lack of spectroscopically determined mass ratio prevent 
an accurate analysis of V49, and for the moment only an approximate photometric 
solution is  possible.  
We made simultaneous fits to $V$ and $B$ light curves from the  
2001-2009 observing seasons using the Phoebe package (Pr\v{s}a \& Zwitter 2005) for a number 
different values of orbital inclination $i$ and mass ratio $q$. The best fit 
was obtained for a semidetached configuration with the primary component 
(i.e. the one eclipsed at phase 0.0) filling its Roche lobe. We derived 
$q=m_{s}/m_{p}=2.0$, $i=78.3$ deg, the ratio of the radii $r_{s}/r_{p} = 0.21$ and 
the ratio of luminosities in $V$ band $L_{s}/L_{p}=23$. Such a configuration
is quite unusual, implying that the light curves are strongly dominated by 
the ellipsoidal variability of the less massive but larger and more luminous 
component. If true, the high luminosity ratio would unfortunately preclude 
a spectroscopic determination of the mass ratio for this interesting system.
 
The binary V66 with $P\approx2.0$ d is a PM member of the cluster. However,
its location 1.4 mag above the main sequence is not consistent with the presumed 
membership. The analysis of template images shows no evidence for any unresolved 
companion which could increase the observed brightness of the binary. On archival 
ACS/HST images only a faint ($V$=23.6 mag) companion is seen at an angular distance 
of 1.3 arcsec. There is still a possibility that V66 is a member of a triple 
hierarchical system. Such a configuration is in fact common among main sequence 
eclipsing binaries with  a period of a few days (Tokovinin et al. 2006). Shallow 
eclipses of similar depth observed in the light curve of V66 are consistent with 
this possibility, although they may be equally well explained by grazing eclipses. 
Spectroscopic observations could clarify this issue.

Contact binaries belonging to the cluster are present only among the blue stragglers. 
These are the variables V39 and V59. The absence of such objects on the main
sequence is remarkable. The analysed sample included 16200  main sequence 
stars with $18.5<V<22.0$. Due to short periods and characteristic sine-like light curves, contact
binaries are very easy to detect even with noisy photometry. A search for
such binaries in NGC 6362 based on our data should be complete down 
to $V\approx22$~mag. The paucity of contact binaries among main 
sequence stars was also observed in our earlier studies of globular clusters
M55, M4 and NGC 6752 (Kaluzny et al. 2010, 2013; Kaluzny \& Thompson 2009). 
Our findings indicate that, at
least in globular clusters, the principal factor enabling contact systems
to form from close but detached binaries is  nuclear evolution: a contact
configuration is achieved once the more massive component starts to expand quickly 
at the turnoff. Apparently,  nuclear evolution is more important in this respect 
than the frequently invoked magnetic breaking; see e.g. Stepien \& Gazeas
(2012) and references therein.

\subsection{Spotted red giants}
Besides V49, there are two other variables belonging to the cluster and 
located on or close to the red giant branch. The light curve of V68 shows 
a sinusoidal modulation with a period of $\sim$18 d. This modulation is coherent 
over the whole duration of our observations. However the average magnitude and amplitude of the 
variability shows seasonal changes. If the observed variations are 
due to binarity then the orbital period would be two times longer than listed 
in Table~1. The light curve of V61 shows an incoherent, roughly sinusoidal 
modulation with a period of $\sim$45 d. In Fig. \ref{fig:curves3} both 
V68 and V61 light curves correspond to the 2001 season.

\subsection{Blue stragglers}
The CMD based on the du Pont data contains 84 
candidate blue stragglers (BS) with $16.0<V<18.7$ and 
$0.05<B-V<0.49$. Of these 70 are 
proper motion members of NGC~6362, 5 are field stars and for 9 objects a
proper motion is not available. The BS lacking a measured proper motion
are located in the very central part of the cluster and are likely 
members. Eight BSs are  variable stars: 
two are contact binaries and six are SX Phe pulsators. The most interesting among the
latter is V38, distinguished by  pulsations with a large 
amplitude (0.63 mag). The fundamental radial mode is dominant with 
$P_1=0.06661582$~d, but the slightly ``diffuse'' light curve indicates that, 
like many other SX Phe stars, V38 is a multiperiodic pulsator. Indeed, we 
found a significant power at $P_2=0.05457972$ d and $P_3=0.04948753$ d.
No variables were detected among candidate BSs 
from the outer part of the cluster found on images from the Swope telescope.
An interesting interloper in the blue stragglers' region is the contact 
binary V56. Using the calibration of Rucinski (2000) we estimated its absolute 
luminosity at $M_{\rm V}=1.87$. At the observed magnitude of $V=17.50$ an apparent
distance modulus is $(m-M)_{V}=15.63$, while for the cluster we have 
$(m-M)_{V}=14.68$ (Harris 1996; 2010 edition). For $A_{V}=0.21$ the variable is located at 
a distance of 12.1 kpc, i.e. 4.3 kpc behind the cluster. Given the galactic 
latitude of $-17.6$~deg, V56 resides about 3.7 kpc above the galactic plane and
is thus an example of a binary blue straggler from the galactic halo.    
\Section{Summary}
 \label{sec:sum}
We conducted an extensive photometric survey of the globular cluster NGC~6362
in a search for variable stars. Twenty five new variables were discovered, and 
multiseasonal light curves were compiled for another 44 variables that had been 
known before. For all variables accurate periods were obtained. One new eclipsing 
binary and two new pulsating stars were found in the blue-straggler region. 
Three of the new detached eclipsing binaries reside in the turnoff region, and 
another two redward of the lower main sequence. Two objects are interesting targets for follow-up observations: V65 at the turnoff
and V49 on the red giant branch.
V65 is a well detached eclipsing binary whose study would tighten constraints on the age 
and distance of the cluster provided by V40 and V41 (Kaluzny et al., 
in preparation). V49 is the first known example of a close red-giant binary 
belonging to a globular cluster. As such, it is a potential source of very valuable 
information concerning the advanced evolutionary stages of low-metallicity stars.

\Acknow
{JK, WN, WP and MR were partly supported by the grant DEC-2012/05/B/ST9/03931
from the Polish National Science Center. 
}

\clearpage
\begin{table}
\footnotesize
 \begin{center}
 \caption{Basic data of NGC~6362 variables identified within the present survey
          \label{tab:vardata}}
 \begin{tabular}{|l|c|c|c|c|c|l|l|c|}
  \hline
 ID & RA & DEC & $V$ & $B-V$ & $A_V$ &Period[d] & Type$^c$ & Mem\\ 
  \hline
V38& 262.93172& -67.04958& 17.02& 0.28 &0.63 &0.06661582(1)  &SX,BS     &Y\\
V39& 263.03523& -67.05412& 17.86& 0.39 &0.20 &0.3632599(1)   &EW,BS     &Y\\
V40& 263.01702& -67.06264& 18.22& 0.55 &0.54 &5.2961749(1)   &EA        &Y\\
V41& 262.89739& -67.06754& 18.77& 0.57 &0.58 &17.888844(4)   &EA        &Y\\
V42& 262.88250& -67.01098& 17.51& 0.69 &0.64 &2.769390(6)    &EA        &N\\
V45& 262.83317& -67.05624& 16.92& 0.80 &0.33 &0.3407063(1)   &EW        &N\\
V46& 263.10395& -67.00871& 17.55& 0.30 &0.05 &0.050634688(5) &SX,BS     &Y\\
V47& 263.05427& -67.04386& 17.35& 0.29 &0.05 &0.052234111(5) &SX,BS     &Y\\
V48& 262.99927& -67.06386& 17.12& 0.33 &0.06 &0.047920021(5) &SX,BS     &Y\\
V49& 263.00796& -67.07682& 14.97& 0.93 &0.36 &32.504(1)      &EB,RG     &Y\\
V53& 263.15424& -67.00576& 19.61& 0.79 &0.25 &0.2826720(1)   &EW        &N\\
V54& 263.22510& -67.09877& 20.41& 0.86 &0.48 &0.4430415(2)   &EA        &Y\\
V55& 263.13284& -67.02132& 17.35& 0.57 &0.05 &6.49353(5)     &Ell       &N\\
V56& 263.13213& -67.03085& 17.50& 0.23 &0.09 &0.5193450(3)   &EW        &N\\
V57& 263.12126& -67.01887& 16.62& 0.99 &0.08 &0.81848(1)$^a$     &Sp?       &N\\
V58& 263.11699& -67.14543& 17.96& 0.95 &0.09 &3.815(1)$^a$       &Sp        &N\\
V59& 263.10289& -67.11148& 17.48& 0.31 &0.03 &0.6601102(4)   &EW/Ell,BS &Y\\
V60& 263.05838& -67.01875& 17.89& 0.92 &0.12 &6.0255(5)$^b$      &Sp        &N\\
V61& 263.07557& -67.06092& 17.23& 0.83 &0.07 &46.35(1)$^b$       &Sp,RG     &Y\\
V62& 263.02428& -67.05223& 19.39& 0.55 &0.33 &16.73212(6)    &EA       &Y\\
V63& 263.01446& -67.13954& 17.80& 0.48 &0.15 &1.890156(3)    &EB        &N\\
V64& 262.99266& -67.06273& 17.06& 0.31 &0.03 &0.050162402(5) &SX,BS     &Y\\
V65& 262.94883& -67.06480& 18.39& 0.55 &0.43 &30.9723(1)     &EA        &Y\\
V66& 262.95013& -67.03277& 19.13& 0.71 &0.20 &1.974206(4)    &EA        &Y\\
V67& 262.93970& -67.07393& 19.55& 0.72 &0.52 &0.25080101(6)  &EW        &N\\
V68& 262.93710& -67.05581& 17.62& 0.77 &0.08 &17.98(1)$^b$       &Sp,RG     &Y\\
V69& 262.93217& -67.02965& 17.14& 1.08 &0.08 &2.0125(1)      &Sp        &N\\
V70& 262.91209& -67.04824& 17.60& 0.49 &0.47 &0.3393452(3)   &EW        &N\\
V71& 262.90240& -67.03735& 19.17& 0.56 &0.30 &11.965385(5)      &EA        &Y\\
V72& 262.87070& -67.04273& 17.61& 0.29 &0.03 &0.0436729(1)   &SX,BS     &Y\\
V73& 262.82041& -67.06011& 20.86& 1.11 &0.82 &0.338903(5)    &EB        &N\\
V74& 262.82326& -66.99927& 18.82& 0.76 &0.17 &0.261080(3)    &EW        &N\\
V75& 262.88797& -67.02960& 18.40& 1.11 &0.10 &1.2290(2)$^a$      &Sp        &N\\
V76& 262.76813& -67.05666& 17.04& 0.74 &0.09 &30.96345(5)      &EA        &N\\
V77& 262.86987& -67.03463& 18.41& 0.89 &0.52 &6.2746(1)      &EA        &-\\
  \hline
 \end{tabular}
\end{center}
{\footnotesize 
$^a$for the 2001 season. $^b$for the 1999 season. 
$^c$Types: EW - contact binary, EB - close eclipsing binary,
EA - detached eclipsing binary, SX - SX Phe star,
Sp - spotted variable, BS- blue straggler, Ell - ellipsoidal variable,
RG - red giant. For V38-V49, which were described by Mazur et al. (1999), 
updated parameters are provided. V53-V77 are new detections.}
\end{table}

\clearpage

\begin{figure}
   \centerline{\includegraphics[width=0.95\textwidth,
               bb = 173 44 828 699, clip]{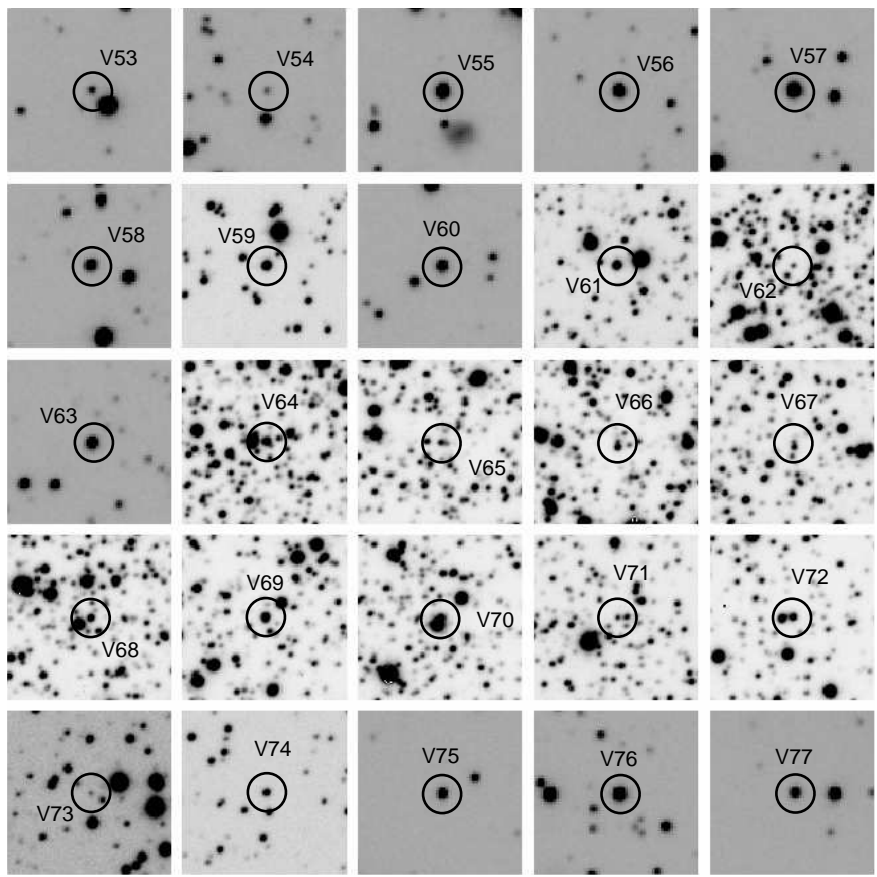}}
   \caption{Finding charts for the newly detected variables V53-77.
    Each chart is 30 arcsec on a side; north is up and east to the left.
    \label{fig:maps}}
\end{figure}

\begin{subfigures}
\begin{figure}
   \centerline{\includegraphics[width=0.95\textwidth,
               bb = 42 25 528 768, clip]{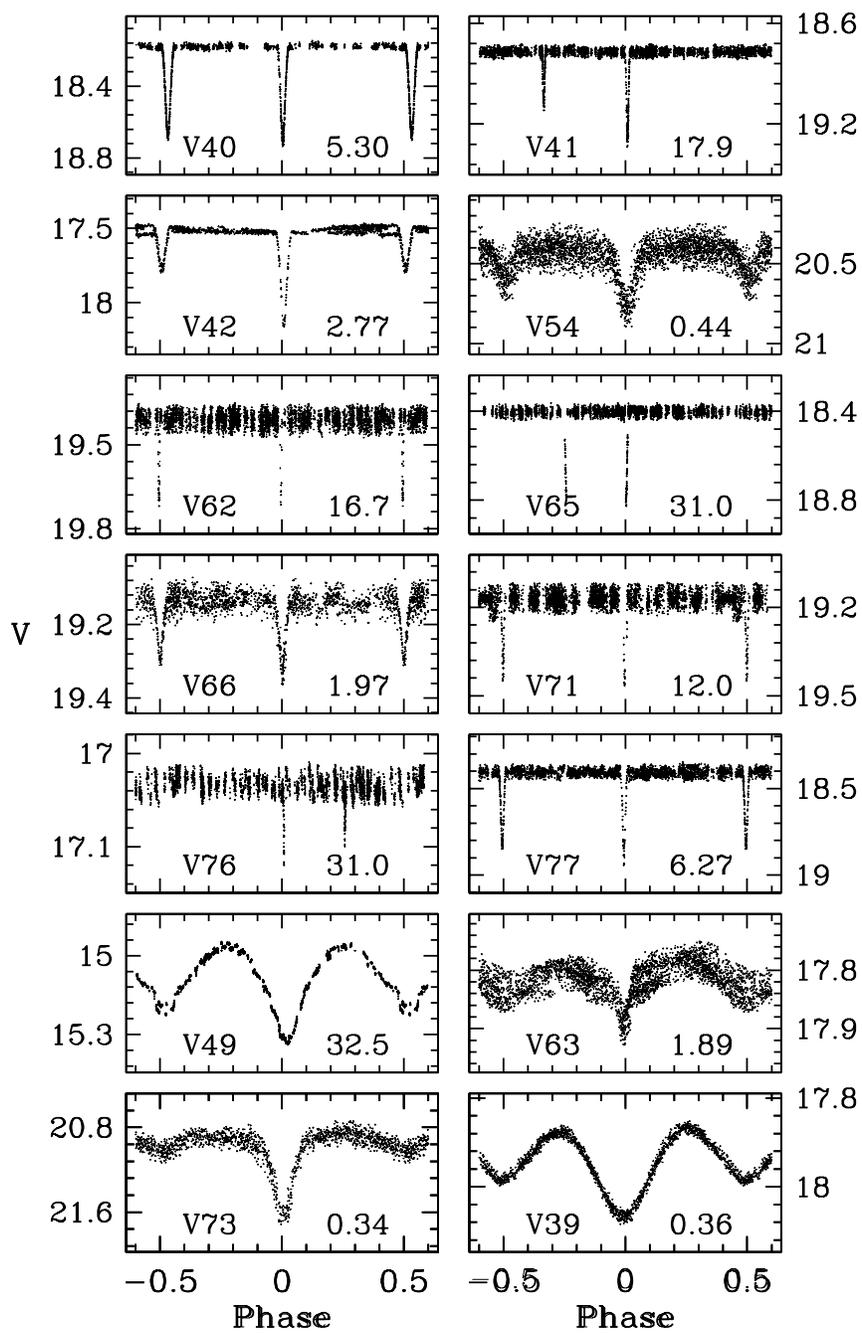}}
   \caption{Phased $V$ curves of variables detected in the field of
    NGC 6362. Inserted labels give star ID and period in days
    \label{fig:curves1}}
\end{figure}

\begin{figure}
   \centerline{\includegraphics[width=0.95\textwidth,
               bb = 42 25 528 768, clip]{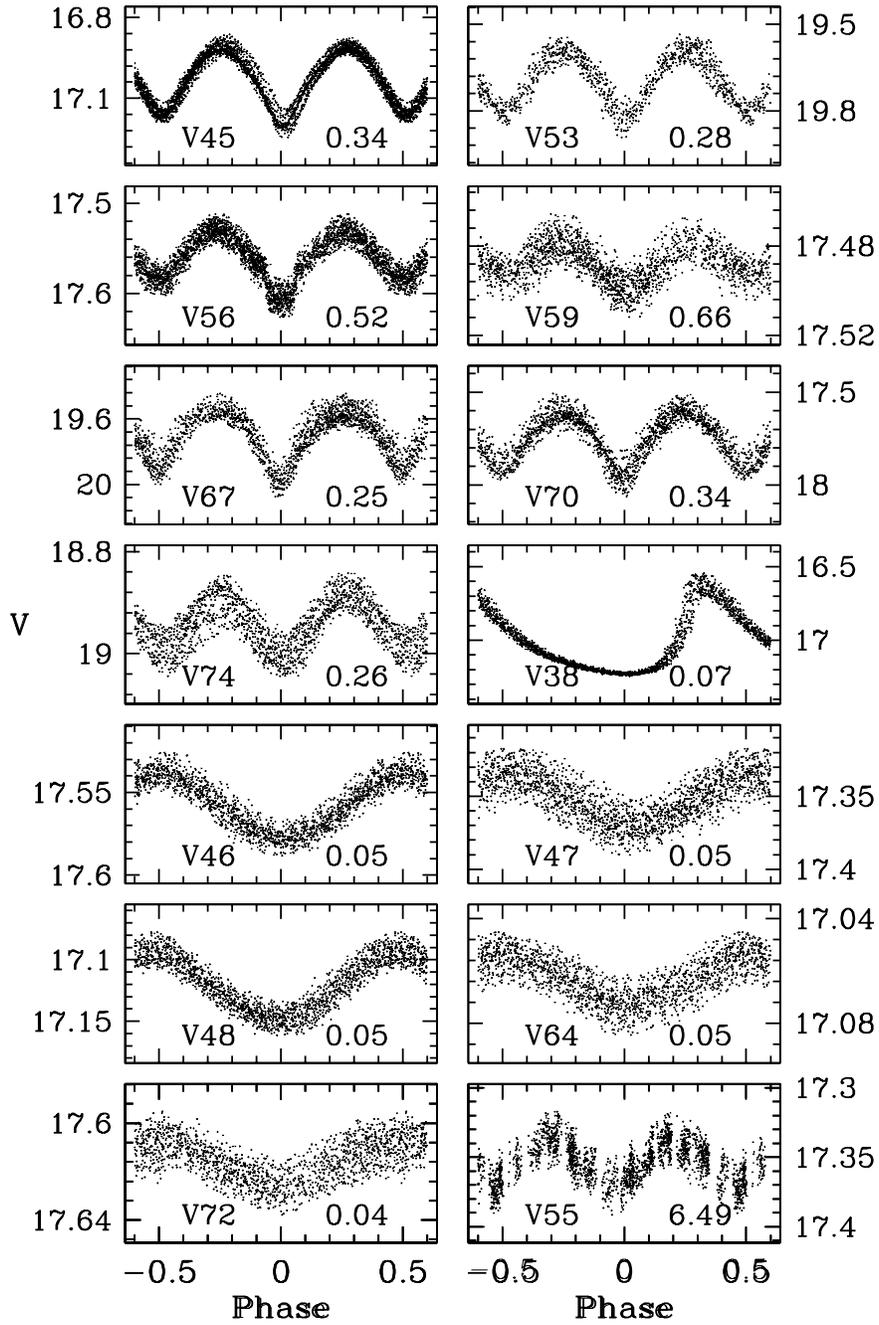}}
   \caption{Continuation of Fig. 4a. The V55 curve shows data from seasons 
    1999-2003. 
    \label{fig:curves2}}
\end{figure}

\begin{figure}
   \centerline{\includegraphics[width=0.95\textwidth,
               bb = 42 328 528 768, clip]{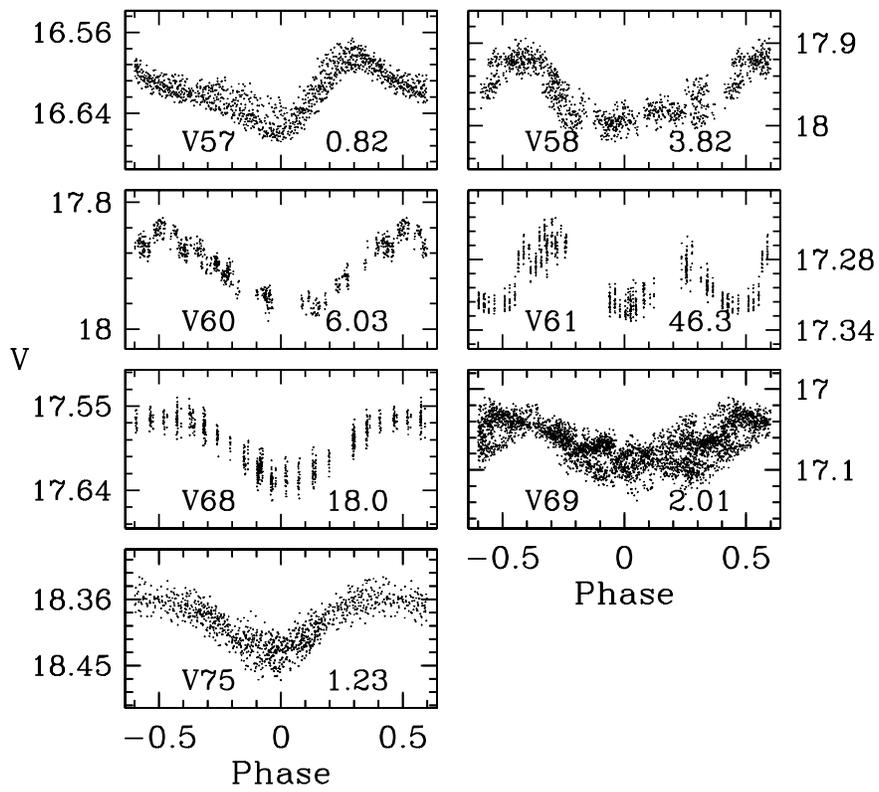}}
   \caption{Continuation of Fig. 4a. Curves of V57, V58 and V75 show data 
    from the 2001 season; curves of V60, V61 and V68 show data from the 1999 season. 
    \label{fig:curves3}}
\end{figure}
\end{subfigures}

\begin{figure}
   \centerline{\includegraphics[width=0.95\textwidth,
               bb = 52  52 562 739, clip]{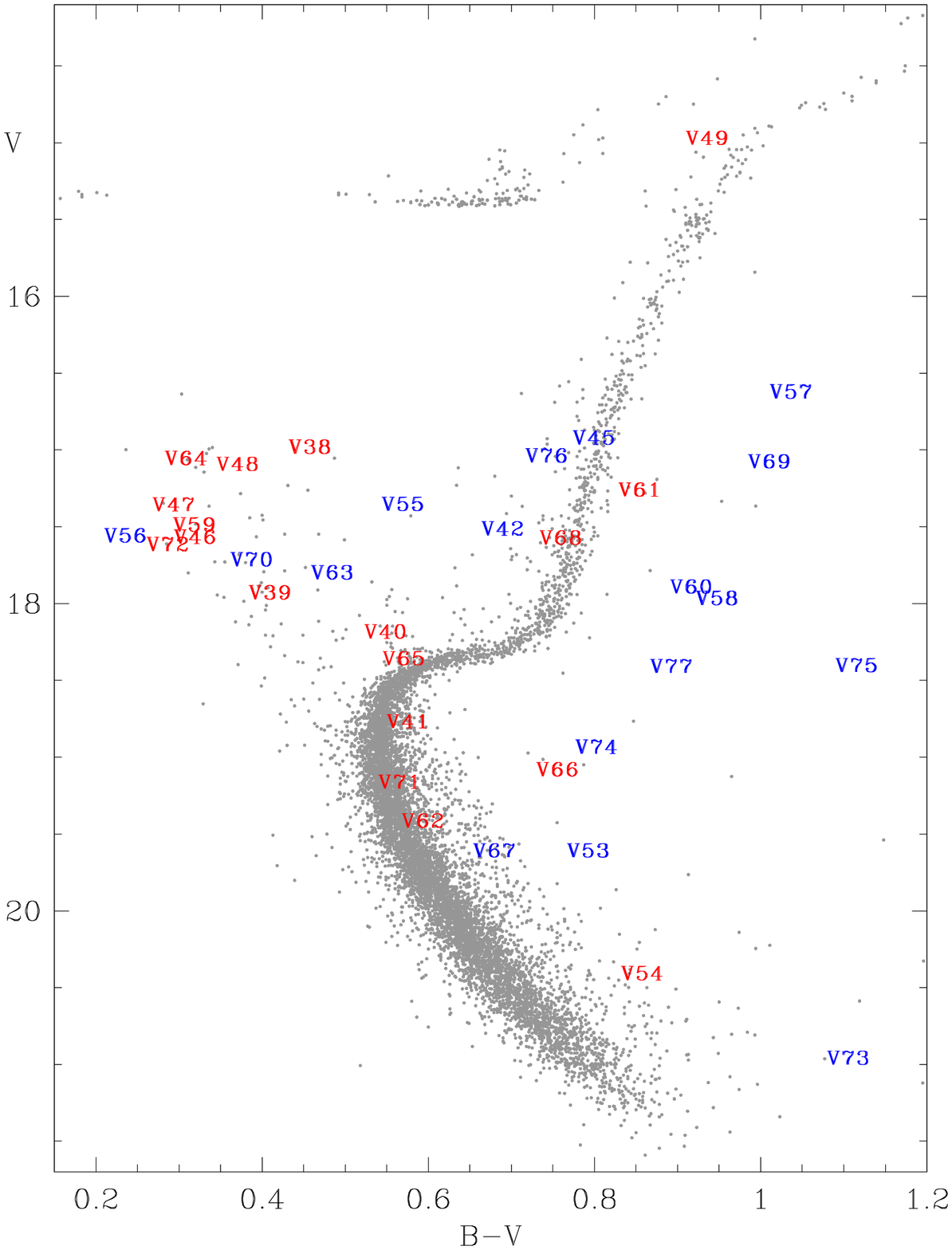}}
   \caption{Color-magnitude diagram for NGC~6362. Points are PM members of the cluster.
    Variables being PM members are marked in red; the remaining ones - in blue. 
    The location of each variable coincides with the center of the corresponding label.
    The CMD is truncated at $V=21.7$ due to the lack of PM information for fainter stars.
    \label{fig:cmdvar}}
\end{figure}

\begin{figure}
   \centerline{\includegraphics[width=0.95\textwidth,
               bb = 44 317 545 715, clip]{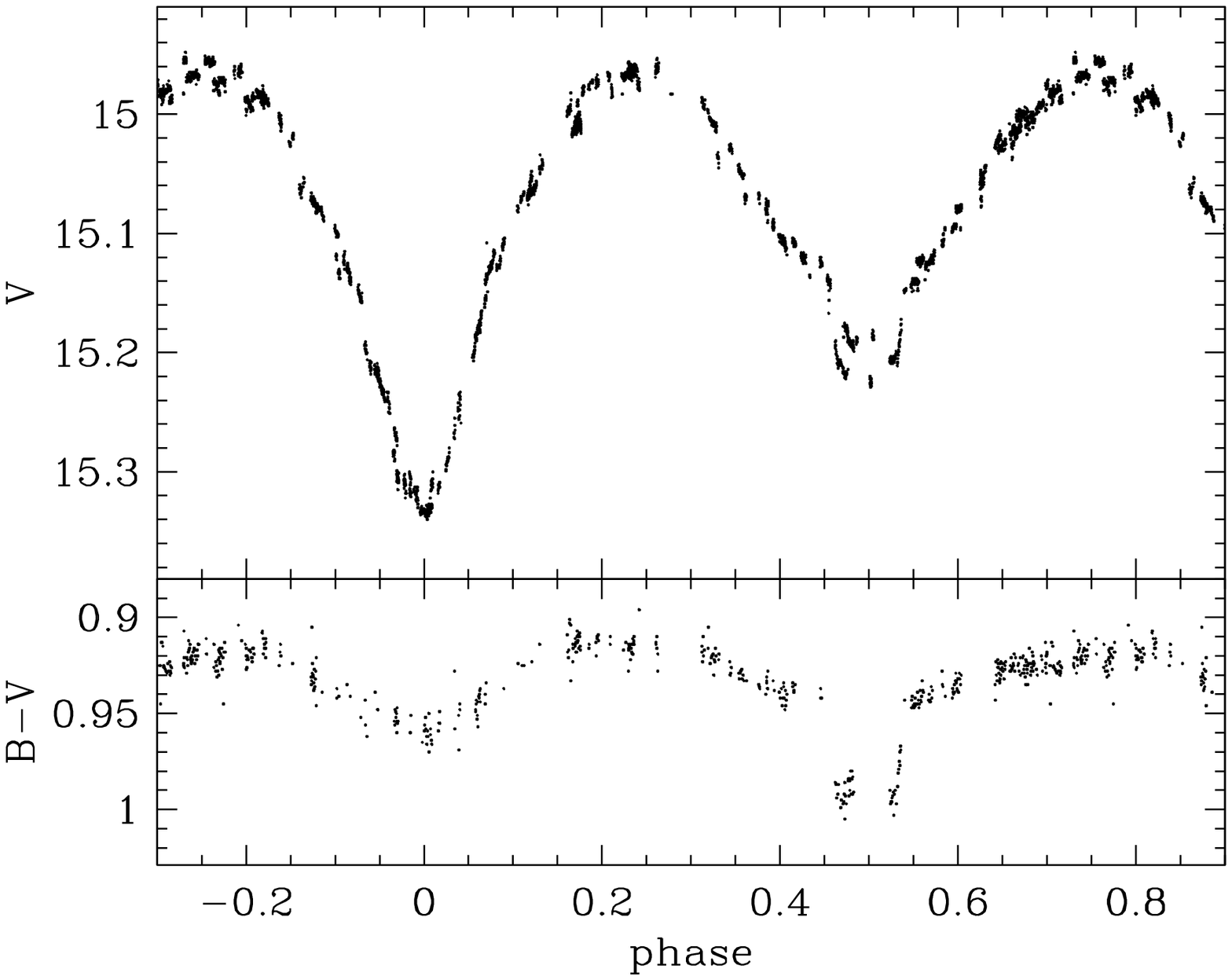}}
   \caption{Light curves of V49.
    \label{fig:V49}}
\end{figure}


\begin{references}

\refitem{Alard, C.}{2000}{A\&AS}{144}{363}	

\refitem{Clement, C. M., Muzzin, A., Dufton, Q., Ponnampalam, T., 
         Wang, J. et al.}{2001}{\AJ}{122}{2587}

\refitem{Harris, W.E.}{1996}{\AJ}{112}{1487}

\refitem{Kaluzny, J., Thompson, I. B.}{2009}{\Acta}{59}{273}

\refitem{Kaluzny, J., Thompson, I. B., Krzeminski, W., Zloczewski, K.}{2010}{\Acta}{60}{246}

\refitem{Kaluzny, J., Thompson, I. B., Krzeminski, W., Preston, G. W., Pych, W.
        et al.}{2005}{Stellar Astrophysics with the World’s Largest Telescopes,
         AIP Conf. Proc.}{752}{70}

\refitem{Kaluzny, J., Thompson, I. B., Rozyczka, M., Krzeminski, W.}{2013}{\Acta}{63}{181}

\refitem{Landolt, A.}{1992} {\AJ}{104}{372}

\refitem{Mathieu, R.D., Meibom, S., Dolan, C. J.}{2004}{\ApJ}{602}{L121}

\refitem{Mazeh, T.}{2008}{EAS Publ. Series}{29}{1}

\refitem{Mazur, B., Kaluzny, J., Krzeminski, W.}{1999}{\MNRAS}{306}{727}

\refitem{Olech, A., Kaluzny, J., Thompson, I. B., Pych, W., Krzeminski, W., 
        Schwarzenberg-Czerny, A.}{2001}{\MNRAS}{321}{421}

\refitem{Pr\v{s}a, A., and Zwitter, T.}{2005}{ApJ}{628}{426}

\refitem{Rucinski, S. M.}{2000}{\AJ}{120}{319}

\refitem{Schwarzenberg-Czerny A.}{ 1996}{\ApJL}{460}{L107}

\refitem{Schwarzenberg-Czerny A., Beaulieu, J.-Ph.}{2006}{\MNRAS}{365}{165}

\refitem{Stepien, K., Gazeas, K.}{2012}{\Acta}{62}{153}

\refitem{Stetson P. B.}{1987}{\PASP}{99}{191}

\refitem{Stetson P. B.}{1990}{\PASP}{102}{932}

\refitem{Tokovinin, A., Thomas, S., Sterzik, M., Udry, S.}{2006}{\AA}{450}{681}

\refitem{Zacharias, N., Finch, C. T., Girard, T. M., Henden, A.,  Bartlett, J. L.
        et al.}{2013}{\AJ}{145}{44}

\refitem{Zloczewski, K., Kaluzny, J., Rozyczka, M., Krzeminski, W., Mazur, B.}
        {2012}{\Acta}{62}{357}
        
\end{references}
\end{document}